\documentstyle[12pt,dina4p]{article}
\newcommand{\be}{\begin{equation}}
\newcommand{\ee}{\end{equation}}
\newcommand{\ba}{\begin{eqnarray}}
\newcommand{\ea}{\end{eqnarray}}
\newcommand{\bc}{\begin{center}}
\newcommand{\ec}{\end{center}}
\newcommand{\vs}{\vspace*{3mm}}
\def\sign{\mbox{\rm sign}}
\def\Tr{\mbox{\rm Tr}}
\def\CS{\mbox{\rm \scriptsize  CS}}
\def\ren{\mbox{\rm \scriptsize  ren}}
\def\bare{\mbox{\rm \scriptsize bare}}
\def\D{\not\!\!D}
\def\DD{\not\!\!\!D}
\begin{document}
\noindent
\begin{titlepage}
\begin{flushright}
DESY 95--111\\
hep-th/9511128
\end{flushright}
  \vspace{1.0in} \LARGE
  \begin{center}
    {\bf Effective Average Action of Chern-Simons Field Theory
     }\\ \vspace{0.3in}
  \end{center}
  \vspace{0.2in} \large
  \begin{center}
    {\rm {M. Reuter} \\
     Deutsches Elektronen-Synchrotron DESY\\
     Notkestrasse 85  \\
     D-22603 Hamburg   \\
     Germany
     }\\ \vspace{0.3in}
  \end{center}
  \vspace{1.5in} \normalsize

\begin{abstract}
The renormalization of the Chern-Simons parameter is investigated by
using an exact and manifestly gauge invariant
evolution equation for the scale-dependent
effective average action.
\end{abstract}

\end{titlepage}

\section*{1. Introduction}

Pure Chern-Simons field theory in 3 dimensions is a fascinating
topic from many points of view. It can be used to give a
path-integral representation of knot and link invariants
\cite{wi} and in order to understand many properties of
2-dimensional conformal field theories \cite{wi,conf}. Being a
topological field theory the model has no propagating degrees of
freedom. In fact, canonical quantization \cite{can} yields a
Hilbert space with only finitely many physical states which can be
related to the conformal blocks of (rational) conformal field
theories. Perturbative covariant quantization
\cite{pir,alv,gia,fal,mar,shif} shows that the theory is not only
renormalizable but even ultraviolet finite. It is
remarkable that despite this high degree of "triviality" the
theory produces nontrivial radiative corrections.
Pisarski and Rao \cite{pir} and Witten
\cite{wi} showed that one-loop effects lead to a
renormalization of the parameter $\kappa$ which multiplies the
Chern-Simons 3-form in the action,
\be
     S_{\CS}[A]=i\kappa \ \frac{g^2}{8\pi} \int d^3\!x
     ~\varepsilon_{\alpha\beta\gamma} \: [A^a_\alpha\,
     \partial_{\beta}\! A^a_\gamma + \frac{1}{3}g
     f^{abc} A^a_{\alpha} A^b_{\beta} A^c_{\gamma}]
\ee
A variety of gauge invariant regularization methods, including
spectral flow arguments based upon the $\eta$-invariant, predict
a finite difference between the bare and the renormalized value
of
$\kappa$:
\be
     \kappa_{\ren}=\kappa_{\bare}+\sign(\kappa)~T(G)
\ee
Here $T(G)$ denotes the value of the quadratic Casimir operator
of the gauge group $G$ in the adjoint representation. It is
normalized such that $T(SU\!(N))=N$. The shift of $\kappa$ has
a natural relation to similar shifts in the Sugawara construction
of 2-dimensional conformal field theories. On the other hand, in
standard renormalization theory a relation of the type (2) is
rather unusual. In a generic renormalizable but not necessarily
finite theory the divergent parts of the counterterms are fixed
by the requirement that the renormalized quantities should be
finite. Their finite parts are not fixed by any general principle
but rather depend on the renormalization scheme. It was argued
that, as there are no ultraviolet divergences in Chern-Simons
theory, there exists a distinguished natural renormalization
scheme which leads to $\kappa_{\ren}=\kappa_{\bare}$ \cite{gia}.
This contradicts the relation (2) favored by conformal field
theory, but it is clear that any argument in favor of one of the
two possibilities must come from considerations which lie outside
the standard framework of renormalized perturbation theory.

\vs

In this paper we investigate Chern-Simons theory along the lines
of the Wilsonian renormalization group approach by using an
exact evolution equation for gauge theories which was
introduced recently \cite{ex,qcd}. It describes the scale
dependence of the effective average action $\Gamma_k$ which can
be thought of as a continuous interpolation between the classical
action $S \equiv \Gamma_{k \rightarrow \infty}$ and the
conventional effective action $\Gamma \equiv \Gamma_{k
\rightarrow 0}$. It depends on the infrared cutoff scale $k$ in
such a way that the functional $\Gamma_k$ evolves out of the
classical action by integrating out only those quantum
fluctuations which have momenta larger than $k$. When $k$ is
lowered from infinity to zero, $\Gamma_k$ follows a certain
trajectory in the space of all actions. This trajectory is a
solution of the exact renormalization group equation \cite{ex}
\ba
     k\frac{d}{dk} \Gamma_k[A,\bar{A}] & = & \frac{1}{2} \Tr
\left[\left(\Gamma_k^{(2)}[A,\bar{A}] +R_k(\Delta[\bar{A}])\right)^{-1}
     k \frac{d}{dk}R_k(\Delta[\bar{A}])\right]
               \\
     && -\Tr\left[\left(-D_{\mu}[A]\, D_{\mu}[\bar{A}] +
     R_k(-D^2(\bar{A}))\right)^{-1} k \frac{d}{dk}
     R_k(-D^2[\bar{A}])\right]    \nonumber
\ea
We use the background gauge fixing technique \cite{abb}.
Therefore $\Gamma_k$ depends on two gauge fields: the usual
classical average field $A^a_\mu$ and the background field
$\bar{A}^a_\mu$.
Eq.(3) has to be solved subject to the initial condition
\be
     \Gamma_{\infty}[A,\bar{A}] =S[A]+\frac{1}{2\alpha} \int
     d^d\!x~
 \left(D^{ab}_{\mu}[\bar{A}]~(A_{\mu}^a-\bar{A}_{\mu}^a)\right)^2
\ee
where the classical action is augmented by the background gauge
fixing term. Furthermore, $\Gamma^{(2)}_k[A,\bar{A}]$ denotes the
matrix of the second functional derivatives of $\Gamma_k$ with
respect to $A$. The function $R_k$ specifies the precise form of
the infrared cutoff. It has to satisfy $\lim_{u \rightarrow 0}
R_k(u)=k^2$, but is arbitrary otherwise. A convenient choice is
\be
     R_k(u)=u~[\exp{(u/k^2)}-1]^{-1}
\ee
but in some cases even a simple constant $R_k=k^2$ is sufficient.
Observable quantities will not depend on the form of $R_k$. A
similar remark applies to the precise form of the operator
$\Delta[\bar{A}] \equiv -D^2 [\bar{A}]+...$ which is essentially
the covariant laplacian, possibly with additional nonminimal
terms \cite{ex,qcd}. The r\^{o}le of $\Delta$ is to distinguish
``high momentum modes" from ``low momentum modes". If one expands
all quantum fluctuations in terms of the eigenmodes of $\Delta$,
then it is the modes with eigenvalues larger than $k^2$ which are
integrated out in $\Gamma_k$. The solution $\Gamma_k[A,\bar{A}]$ of (3)
with (4) is gauge invariant under simultaneous gauge
transformations of $A$ and $\bar{A}$. In practice solutions can
be found by truncating the space of actions to a finite
dimensional subspace. If one makes an ansatz for $\Gamma_k$ which
contains only finitely many parameters (depending on $k$) and
inserts it into (3), the functional differential equation
reduces to a set of coupled ordinary differential equations for
the parameter functions \cite{qcd,ahm}.

The effective average
action $\Gamma_k$ is closely related to a continuum version of
the block-spin action of lattice systems\footnote{Also in
ref.\cite{shif} a version of the Wilsonian effective action was
used.}. Block-spin transformations can be iterated, and when we
have already constructed $\Gamma_{k_1}$ at a certain scale $k_1$
we may view $\Gamma_{k_1}$ as the ``classical" action for the next
step of the iteration, in which an integral over
$\exp{(-\Gamma_{k_1})}$ has to be performed. If we now apply this
machinery to Chern-Simons field theory and try to understand the
shift (2) from a renormalization group point of view, we are
immediately confronted with the following puzzle. Because
$S_{\CS}$ is not invariant under large gauge transformations,
$\exp{(-S_{\CS})}$ is single valued only if $\kappa \in {\bf Z}$. In
the renormalization group language $\kappa_{\bare}$ has to be
identified with $\kappa(k=\infty)$ and $\kappa_{\ren}$ with
$\kappa(k=0)$, where $\kappa=\kappa(k)$ is the scale-dependent
prefactor of the Chern-Simons term. If there is a smooth
interpolation between $\kappa(\infty)$ and $\kappa(0)$ a
nontrivial shift (2) implies that there are intermediate scales at
which $\kappa$ cannot be integer. Hence it seems that there
should be an inconsistency if we try to do the next blockspin
transformation starting from a scale $k_1$ where $\kappa$ is
non-integer, because we would have to integrate over a
multivalued ``Boltzmann factor" $\exp{(-\Gamma_{k_1})}$. Thus
one is led to believe that any sensible solution to the
evolution equation should have $\kappa_{\ren}=\kappa_{\bare}$.
In the following we shall see that this ``no go theorem" is
actually wrong: a non-zero shift is not in contradiction with a
well-defined (albeit somewhat unusual) renormalization group
trajectory.

\section*{2. Truncating the Evolution Equation}

Let us try to solve the initial value problem (3) with (4) for
the classical Chern-Simons action (1). We work on flat euclidean
space and allow for an arbitrary semi-simple, compact gauge group
$G$. Our strategy for finding solutions of the evolution equation
is to restrict the infinite dimensional space of all actions to a
finite dimensional subspace by means of an appropriate ansatz for
$\Gamma_k$. In the case at hand the essential physics is captured
by a $\Gamma_k$ of the form
\ba
     \Gamma_k[A,N,\bar{A}] &=&
     i\kappa(k)~\frac{g^2}{4\pi}~I[A]+
     \kappa(k)~\frac{g^2}{8\pi}    \int d^3\!x\,\Bigl\{iN^a
     D_{\mu}^{ab}[\bar{A}]\, (A^b_{\mu}-\bar{A}^b_{\mu})
         \\   \nonumber
     &&-i(A^a_{\mu}-\bar{A}^a_{\mu})D^{ab}_{\mu}[\bar{A}]~N^b
      +\alpha\, \kappa(k)
     \frac{g^2}{4\pi}N^aN^a\Bigr\}
\ea
with
\be
     I[A] \equiv \frac{1}{2} \int
     d^3\!x~\varepsilon_{\alpha\beta\gamma}
     ~[A^a_{\alpha}\,\partial_\beta\! A^a_\gamma +
     \frac{1}{3}g f^{abc}
     A^a_{\alpha}A^b_{\beta}A^c_{\gamma}]
\ee
The first term on the RHS of (6) is the Chern-Simons action, but
with a scale-dependent prefactor. In the second term we
introduced an auxiliary field $N^a(x)$ in order to linearize the
gauge fixing term. By eliminating $N^a$ one recovers the
classical, $k$-independent background gauge fixing term
     $\frac{1}{2\alpha}(D_{\mu}[\bar{A}](A_{\mu}-\bar{A}_{\mu}))^2$.
In principle also the gauge fixing term could change its form
during the evolution, but this effect is neglected here.
The ansatz (6) is motivated by the success of similar truncations
in 4 dimensions. Apart from the gauge fixing term we keep only
the dimension-3 operator and neglect all terms which are
``irrelevant'' according to their canonical dimension.
It was demonstrated already that
in QCD
\cite{ex,qcd} and in the abelian Higgs model \cite{ahm} the
approximation of keeping only the relevant and the marginal terms
can lead to rather accurate results which go well beyond a one-loop
calculation.

\vs

For $k\rightarrow \infty$, and upon eliminating $N^a$, the ansatz
(6) reduces to (4) with the identification
     $\kappa(\infty) \equiv \kappa_{\bare}$.
We shall insert (6) into the evolution equation and from the
solution for the function $\kappa(k)$ we shall be able to
determine the renormalized parameter
     $\kappa(0) \equiv \kappa_{\ren}$.
We have to project the traces on the RHS of (3) on the subspace
spanned by the truncation (6). In practice this means that we have
to extract only the term proportional to $I[A]$ and to compare
the coefficients of $I[A]$ on both sides of the equation. In the
formalism with the auxiliary field $N^a\!,$ $\Gamma^{(2)}_k$ in (3)
denotes the matrix of second functional derivatives with respect
to both $A^a_\mu$ and $N,$ but with $\bar{A}^a_\mu$ fixed
\cite{ex}. As we are only interested in the coefficient of
$I[A]$, it is computationally advantageous to set $\bar{A}=A$
after the derivatives have been performed. Then the second
variation of (6) becomes
 \ba
     \delta^2\Gamma_k[A,N,A]&= & i\kappa(k) \frac{g^2}{4\pi}
     \int d^3\!x~\Big\{\delta\! A^a_{\mu}\, \varepsilon_{\mu \nu
     \alpha} D^{ab}_{\alpha}\, \delta\! A^b_{\nu}+\delta\!
     N^aD^{ab}_{\mu}\,\delta\! A^b_{\mu}   \nonumber \\
                                      && -\delta\!
     A^a_{\mu}D^{ab}_{\mu}N^b\Big\}
     +\alpha \ (\kappa(k) \frac{g^2}{4\pi})^2 \int
     d^3x~\delta\! N^a~\delta\! N^a
 \ea
In order to facilitate the calculations we introduce three
4$\times$4 matrices $\gamma_\mu$ with matrix elements
$(\gamma_\mu)_{mn}$, $m$=($\mu$,4)=1,...,4, etc., in the
following way\cite{shif}:
 \ba
 (\gamma_\mu)_{\alpha\beta}=\varepsilon_{\alpha\mu\beta},
  &
     (\gamma_\mu)_{4\alpha}=-(\gamma_\mu)_{\alpha 4}
     =\delta_{\mu\alpha}
     \nonumber \\
     (\gamma_\mu)_{44}=0 \hspace*{8mm} &
 \ea
If we combine the gauge field fluctuation and the auxiliary field
into a 4-component object $\Psi^a_m \equiv (\delta
A^a_{\mu},\delta N^a)$ and choose the gauge $\alpha=0$, then
(8) assumes the form
\be
     \delta^2\Gamma_k[A,N,A] = i\kappa(k) \frac{g^2}{4\pi}
     \int d^3\!x ~\Psi^a_m(\gamma_\mu)_{mn} D^{ab}_{\mu}
     \Psi^b_n
\ee
so that in matrix notation
\be
     \Gamma^{(2)}_k=i\kappa(k) \  \frac{g^2}{4\pi} \D
\ee
Clearly $\D \equiv \gamma_{\mu}D_{\mu}$ is reminiscent of a Dirac
operator. In fact, the algebra of the $\gamma$-matrices is similar
to the one of the Pauli matrices:
$\gamma_{\mu}\gamma_{\nu}=-\delta_{\mu\nu}
+\varepsilon_{\mu\nu\alpha}\gamma_{\alpha}$.
Because $\gamma^+_{\mu}=-\gamma_{\mu}, \ \ \D$ is hermitian. Its
square reads
\be
     \D^2=-D^2-ig\ ^*\!F_\mu\gamma_{\mu}
\ee
where
     $^*\!F_{\mu} \equiv \frac{1}{2}
     \varepsilon_{\mu\alpha\beta}F_{\alpha\beta}$
is the dual of the field strength tensor. (In equations such as
(11) and (12) $A_\mu$ and $F_{\mu\nu}$ are matrices in the
adjoint representation.) Because $\D^2$ is `almost' equal to the
covariant laplacian, it is the natural candidate for the cutoff
operator $\Delta$. With this choice the evolution equation (3)
reads at $\bar{A}=A$:
\ba
     ic ~k \frac{d}{dk}\kappa(k)~I[A] &=& \frac{1}{2} \Tr
 \left[\left(ic\kappa \D+R_k(\D^2)\right)^{-1} k \frac{d}{dk}R_k(\D^2)%
\right] \nonumber   \\  &&-  \Tr
\left[\left(-D^2+R_k(-D^2)\right)^{-1} k \frac{d}{dk}R_k(-D^2)%
 \right]
\ea
Here $c \equiv g^2/4\pi$. The equality sign in (13) is to be
understood in the sense that the term $\sim i I[A]$ has to be
extracted from the RHS and all other terms have to be discarded.
In particular, the second trace on the RHS of (13) is manifestly
real, so it cannot match the purely imaginary $i I[A]$ and can be
omitted therefore. For the same reason we may replace the first
trace by $i$ times its imaginary part:
\be
     k \frac{d}{dk} \kappa(k)\, I[A] = - \frac{1}{2}
     \kappa(k)\,\Tr\left[\, \D\ \left(c^2 \kappa^2%
     \D^2 +R^2_k(\D^2)\right)^{-1}
     k \frac{d}{dk}R_k(\D^2)\right]
     +\cdots
\ee
The trace in (14) involves an integration over spacetime, a
summation over adjoint group indices, and a ``Dirac trace". We
shall evaluate it explicitly in the next section. Before turning
to that let us first look at the general structure of eq.(14). In
terms of the (real) eigenvalues $\lambda$ of $\D$ eq.(14) reads
\be
     \frac{d\kappa(k)}{dk^2}~I[A]=-\frac{1}{2}\kappa(k)
     \sum_{\lambda}\frac{\lambda}{c^2\kappa^2(k)
     \lambda^2+R^2_k(\lambda^2)} \cdot
     \frac{dR_k(\lambda^2)}{dk^2}
\ee
where we switched from $k$ to $k^2$ as the independent variable.
We observe that the sum in (15) is related to a regularized form
of the spectral asymmetry of $\DD$. We emphasize at this point
that the evolution equation (3), and therefore also (15), is
well-defined, both in the infrared and the ultraviolet, without
any further regularization. If one employs a cutoff function
$R_k(u)$ which vanishes exponentially fast for $u \rightarrow
\infty$ (such as (5) for example) only eigenvalues of $\Delta$ in
a small neighborhood of $\lambda \approx k$ contribute
significantly to the trace \cite{ex}.

\vs

An approximate solution for $\kappa(k)$ can be obtained by
integrating both sides of eq.(15) from a low scale $k^2_0$ to a
higher scale $\Lambda^2$ and approximating $\kappa(k) \simeq
\kappa(k_0)$ on the RHS. (In more conventional theories \cite{ahm}
this type of approximation amounts to neglecting anomalous
dimensions.) This yields
\be
     [\kappa(k_0)-\kappa(\Lambda)]~I[A] = \frac{1}{2}
     \kappa(k_0)
     \sum_{\lambda}
                 \int^{\Lambda^2}_{k_0^2} dk^2
  \ \frac{dR_k(\lambda^2)}{dk^2} \cdot
     \frac{\lambda}{c^2\kappa^2(k_0)
     \lambda^2+R^2_k(\lambda^2)}
\ee
Upon using $R_k$ as the variable of integration one arrives at
\be
     [\kappa(k_0)-\kappa(\Lambda)]~I[A] =
     \frac{1}{2c}\ \sign(\kappa(k_0)) \sum_{\lambda}~
     \sign(\lambda)\, G(\lambda;k_0,\Lambda)
\ee
with
\be
     G(\lambda;k_0,\Lambda) \equiv \arctan\left[
     c\,|\kappa(k_0)\lambda|\,
     \frac{R_\Lambda(\lambda^2)-R_{k_0}(\lambda^2)}
     {c^2\kappa(k_0)^2\lambda^2 +
     R_{\Lambda}(\lambda^2)~R_{k_0}(\lambda^2)} \right]
\ee
Recalling the properties of $R_k$ we see that in the spectral
sum (17) the contributions of eigenvalues $|\lambda| \ll k_0$
and $|\lambda|\gg\Lambda$ are strongly suppressed, and only the
eigenvalues with $k_0 < |\lambda| < \Lambda$ contribute
effectively. Ultimately we would like to perform the limits $k_0
\rightarrow 0$ and $\Lambda \rightarrow \infty$. In this case the
sum over $\lambda$ remains without IR and UV regularization. This
means that if we want to formally perform the limits $k_0
\rightarrow 0$ and $\Lambda \rightarrow \infty$ in eq.(17), we
have to introduce an alternative regulator. In order to make
contact with the standard spectral flow argument \cite{wi} let us
briefly describe this procedure. We avoid IR divergences by
putting the system in a finite volume and imposing boundary
conditions such that there are no zero modes. In the UV we
regularize with a zeta-function-type convergence factor
$|\lambda/\mu|^{-s}$ where $\mu$ is an arbitrary mass
parameter. Thus the spectral sum becomes
\be
     \lim_{s \rightarrow 0}\  \sum_{\lambda} \sign(\lambda)\,
     \left|\frac{\lambda}{\mu}\right|^{-s} G(\lambda;
     k_0,\Lambda)
\ee
Now we interchange the limits $k_0 \rightarrow 0$, $\Lambda
\rightarrow \infty$ and $s \rightarrow 0$. By construction, only
finite $(|\lambda| \leq \mu)$ and nonzero eigenvalues
contribute in (19). For such $\lambda$'s we have
$G(\lambda; 0,\infty)=\pi/2$ irrespective of the precise form of
$R_k$. Therefore (17) becomes
\be
     [\kappa(0)-\kappa(\infty)]~I[A]=
     \frac{2\pi^2}{g^2}~\sign(\kappa(0)) ~\eta[A]
\ee
where $\eta[A] \equiv \lim_{s \rightarrow 0} \frac{1}{2}
\sum_{\lambda} \sign(\lambda)~|\lambda/\mu|^{-s} $ is the
eta-invariant. If we insert the known result \cite{wi}
$\eta[A]=(g^2/2\pi^2)~T(G)~I[A]$ we find that in agreement with
eq.(2)
\be
     \kappa(0)=\kappa(\infty)+\sign(\kappa(0))~T(G)
\ee
We see that at least at the formal level the function $R_k$ has
dropped out of the calculation. In this sense the shift of the
parameter $\kappa$ is universal: it does not depend on the form
of the IR cutoff.

\section*{3. Explicit Calculation}

Next we turn to the evaluation of the trace in eq.(14). The
derivation in this section does not rely on formal manipulations
of spectral sums, and it will keep the full $k$-dependence of
$\kappa$ on the RHS. It is precisely this $\kappa(k)$-dependence
on the RHS of the evolution equation which implements the
``renormalization group improvement" \cite{ex,qcd}. To start with
we use the constant cutoff $R_k=k^2$ for which eq.(14) assumes
the form\footnote{Even with $R_k=k^2$ there are no convergence
problems for $\lambda \rightarrow \infty$ in eq.(15). The
extraction of the term $\sim I[A]$ from the spectral sum involves
derivatives which improve the convergence, see eq.(25) below. }
\be
     \frac{d}{dk^2}\kappa(k)~I[A]
=-\frac{1}{2c^2\kappa(k)}%
 \,\Tr\left[\, \D\left(\D^2+l(k)^2\right)^{-1}\right]
\ee
where
\be
     l(k) \equiv \frac{k^2}{c~|\kappa(k)|}
\ee
(Note that in 3 dimensions $c \equiv g^2/4\pi$ and hence also $l$
has the dimension of a mass.) Our strategy is to extract from the
trace the term quadratic in $A$ and linear in the external
momentum, and to equate the coefficients of the $A\,\partial \!
A$-terms on both sides. (Using the $A^3$-term instead leads to
the same answer.) Using
$tr(\gamma_{\alpha}\gamma_{\mu}\gamma_{\nu})
=-4\varepsilon_{\alpha\mu\nu}$, \ \
$f^{acd}f^{bcd}=T(G)\,\delta^{ab}$ and similar identities one
obtains after some algebra
\be
     \frac{d\kappa(k)}{dk^2} \int d^3\!x
     ~\varepsilon_{\alpha\beta\gamma}
     ~A^a_{\alpha}\,\partial_{\beta}\!A^a_{\gamma}=
     -\frac{g^2T(G)}{c^2\kappa(k)} \int d^3\!x\,
     \varepsilon_{\alpha\beta\gamma}\,
     A^a_{\alpha}\Pi_k(-\partial^2)
     \partial_{\beta}\!A^a_{\gamma}+O(A^3)
\ee
The function $\Pi_k$ is given by the Feynman parameter integral
\be
     \Pi_k(q^2)=8 \int_0^1 dx~x(1-x) \int
     \frac{d^3p}{(2\pi)^3} \,
     \frac{q^2}{[p^2+l^2+x(1-x)q^2]^3}
\ee
Expanding $\Pi_k(-\partial^2)= \Pi_k(0)-\Pi'_k(0)\partial^2+
...$, we see that only for the term with $\Pi_k(0)$ the number of
derivatives on both sides of eq.(24) coincides. Therefore one
concludes that
\be
     \frac{d\kappa(k)}{dk^2}= - \frac{g^2
     T(G)}{c^2\kappa(k)} \  \Pi_k(0)
\ee
where $\Pi_k(0)$ depends on $\kappa(k)$ via (23). Equation (26) is
the renormalization group equation for $\kappa(k)$ which we
wanted to derive. Formally it is similar to the evolution
equations which we derived for QCD\cite{ex} and for the abelian Higgs
model \cite{ahm}. The very special features of Chern-Simons
theory, reflecting its topological character, become obvious when
we give a closer look to the function $\Pi_k(q^2)$. Assume we fix
a non-zero value of $k$ $(l\neq 0)$ and let $q^2 \rightarrow 0$
in (25). Because the $l^2$-term prevents the $p$-integral from
becoming IR divergent, we may set $q^2=0$ in the denominator, and
we conclude that the integral vanishes $\sim q^2$. This means
that the RHS of (26) is zero and
that $\kappa(k)$ keeps the same value for all strictly positive
values of $k$ . One might be tempted to take this result
as a confirmation of the ``no-go theorem" mentioned in the
introduction and to conclude that $\kappa_{\ren}=\kappa_{\bare}$.
This is premature however because $\Pi_k(0)$ really vanishes
only for $k>0$. If we set $l=0$ in (25) we cannot conclude
anymore that $\Pi_k \sim q^2$, because in the region $p^2
\rightarrow 0$ the term $x(1-x)q^2$ provides the only IR cutoff
and may not be set to zero in a naive way. In fact, $\Pi_k(0)$
has a $\delta$-function-like peak at $k=0$. To see this, we first
perform the integrals in (25):
\be
     \Pi_k(q^2)=\frac{1}{\pi}\left[ \frac{1}{2|q|} \arctan
     \left(\frac{|q|}{2|l|}\right)-\frac{|l|}{q^2+4l^2}
     \right]
\ee
As $q^2$ approaches zero, this function develops an increasingly
sharp maximum at $l=0$. Integrating (27) against a smooth test
function $\Phi(l)$ it is easy to verify that
\be
     \lim_{q^2 \rightarrow 0} \int_0^{\infty} dl~
     \Phi(l)~\Pi_k(q^2) = \frac{1}{4\pi} \Phi(0)
\ee
This means that on the space of even test functions $\lim_{q^2
\rightarrow 0}\Pi_k(q^2)=\delta(l)/2\pi$. Even though the value
of $\kappa(k)$ does not change during almost the whole evolution
from $k=\infty$ down to very small scales, it performs a finite
jump in the very last moment of the evolution, just before
reaching $k=0$. This jump can be calculated in a well-defined
manner by integrating (26) from $k^2=0$ to $k^2=\infty$:
\be
     \kappa(0)-\kappa(\infty)= 4\pi~T(G)~\lim_{q^2
     \rightarrow 0} \int_0^{\infty} dl~\sign(\kappa(l))
     \cdot \left[1-c \, l \frac{d}{dk^2} |\kappa(k)|\right]^{-1}
     \Pi_k(q^2)
\ee
The term $\sim d|\kappa|/dk^2$ is a Jacobian factor which is due
to the fact that $l$ depends on $\kappa(k)$. This factor is the only
remnant of the $\kappa(k)$-dependence of the RHS of the evolution
equation. We mentioned already that, in more conventional
theories, this dependence of the RHS on the running couplings is
the origin of the renormalization group improvement. Chern-Simons
theory is special also in this respect. If we use (28) in (29),
$l~d|\kappa|/dk^2$ is set to zero and we find
\be
     \kappa(0)=\kappa(\infty)+\sign(\kappa(0))~T(G),
\ee
which is precisely the 1-loop result. It is straightforward to
check that the shift (30) is independent of the choice for $R_k$. For
a generic cutoff the momentum space integral (25) becomes more
complicated and depends on $R_k$ nontrivially. Nevertheless, by
an argument similar to the one following eq.(16) the relation
(30) can be seen to hold for any $R_k$.

\section*{4. Conclusion}

We used an exact and manifestly gauge invariant evolution
equation in order to study the renormalization of the
Chern-Simons parameter. The method of truncating the space of
actions allows us to obtain nonperturbative solutions which
require neither an expansion in the number of loops nor in the
gauge coupling. The approximation involved here is that
during the evolution the mixing of the Chern-Simons term with other
operators is neglected. This approach has been tested already in
the abelian Higgs model \cite{ahm} and in QCD\cite{ex,qcd}. The
results obtained for Chern-Simons theory are strikingly
different in at least two respects.

\vs

Like $\kappa$, also the gauge coupling in QCD$_4$, for instance,
is a universal quantity. Its running is governed by a
$R_k$-independent $\beta$-function which leads to a logarithmic
dependence on the scale $k$. The Chern-Simons parameter $\kappa$,
on the other hand, does not run at all between $k=\infty$ and any
infinitesimally small value of $k$. Only at the very end of the
evolution, when $k$ is very close to zero, $\kappa$ jumps by a
universal, unambiguously calculable amount $\pm T(G)$. Though
surprising in comparison with non-topological theories, this
feature is precisely what one would expect if one recalls the
topological origin of a non-vanishing $\eta$-invariant \cite{wi}.
If $\eta[A]\neq 0$ for a fixed gauge field $A$,
some of the low lying eigenvalues of $\D[A]$ must have
crossed zero during the interpolation from $A=0$ to $A$. However,
this spectral flow involves only that part of the spectrum which,
in the infinite volume limit, is infinitesimally close to zero.
It is gratifying to see that even without an artificial
discretization of the spectrum (by a finite volume) the spectral
flow is correctly described by the evolution equation. A jump in
$\kappa$, rather than a continuous evolution, resolves the puzzle
mentioned in the introduction: at $k>0$ the iterated block-spin
transformations are all well-defined, but their limit is
nontrivial. It is also remarkable that the evolution equation by
itself is well-defined even for noninteger $\kappa$. The
quantization condition follows only if we require the limit
$\lim_{k\rightarrow 0} \exp(-\Gamma_k)$ to be a single-valued
functional\footnote{A similar phenomenon occurs in stochastic
quantization \cite{sto}.}.

\vs

The second unusual feature of Chern-Simons theory is the absence
of any renormalization group improvement beyond the 1-loop
result. This situation has to be contrasted with the running of
$g$ in QCD$_4$, for instance where a truncation similar to the
one used here leads to a nonperturbative $\beta$-function
involving arbitrarily high powers of $g$. We emphasize that our
exact evolution equation with the truncation (6) potentially goes
far beyond a 1-loop calculation. It is quite remarkable therefore
that in Chern-Simons theory all higher contributions vanish. It
is not possible to translate such a ``nonrenormalization theorem"
for a given truncation into a statement about the
nonrenormalization at a given number of loops. Nevertheless, our
results point in the same direction as ref.~\cite{gia} where the
absence of 2-loop corrections was proven. As there are
gauge-invariant regularizations which do not produce the shift
(2) \cite{pro} it remains an open questions whether
more complicated truncations could modify the above picture.

\noindent
Acknowledgement: I would like to thank E.Gozzi and C.Wetterich for
helpful discussions.

\end{document}